\documentclass[epj]{svjour}
\usepackage{graphicx}
\usepackage{color}
\begin{document}
\title{Promotion of cooperation on networks? The myopic best response case}
\author{Carlos P.\ Roca\inst{1,2} \and Jos\'e A.\ Cuesta\inst{1} \and Angel S\'anchez\inst{1,3,4}
}                     
%
%
\institute{Grupo Interdisciplinar de Sistemas Complejos (GISC), Departamento de
Matem\'aticas, Universidad Carlos III de Madrid, Spain
\and Chair of Sociology, in particular of Modeling and Simulation, ETH Zurich,
Switzerland \and Instituto de Ciencias Matem\'aticas CSIC-UAM-UC3M-UCM, Madrid, Spain \and Instituto de Biocomputaci\'on y F\'\i sica de Sistemas
Complejos (BIFI), Universidad de Zaragoza, Spain}
\date{Received: date / Revised version: date}
%
\abstract{ We address the issue of the effects of considering a network of
contacts on the emergence of cooperation on social dilemmas under myopic best response
dynamics. We begin by summarizing the main features observed under less
intellectually demanding dynamics, pointing out their most relevant general
characteristics. Subsequently we focus on the new framework of best response.
By means of an extensive numerical simulation program we show that, contrary to
the rest of dynamics considered so far, best response is largely unaffected by
the underlying network, which implies that, in most cases, no promotion of
cooperation is found with this dynamics. We do find, however, nontrivial
results differing from the well-mixed population in the case of coordination
games on lattices, which we explain in terms of the formation of spatial
clusters and the conditions for their advancement, subsequently discussing
their relevance to other networks.
} 
\maketitle
\section{Introduction}
\label{intro}

The origin and sustainability of cooperation in animal and human societies is
a long-standing puzzle whose importance cannot be overstated \cite{Pennisi:2005}.
Since the pioneering works by Hamilton \cite{Hamilton:1964,Axelrod:1981},
and the introduction of the theoretical setup of evolutionary game theory
\cite{Maynard-Smith:1982},
a number of reasons have been advanced as
possible explanations for the ubiquity and robustness of cooperative behavior
\cite{Nowak:2006} (see also \cite{Roca:2006}).
Among these proposals, network reciprocity, or the existence
of a (possibly social) network of contacts that governs the individuals a particular
one interacts with, has received much attention in the last two decades. This
specific line of research started with a seminal work by Nowak and May
\cite{Nowak:1992}, who studied the Prisoner's Dilemma (PD) \cite{Rapoport:1965}
game on a
square lattice, finding evidence for substantial amounts of cooperation in
parameter regions where defection was the only possible outcome in a
well-mixed population (i.e., when every individual interacts with every other one).
Subsequent work has explored many other choices for the network as well as
other dynamical rules for the update of strategies in the game, giving rise to
a considerable amount of work \cite{Szabo:2007} which, however, yielded
quite a few contradictory results and no global picture of the observed
phenomenology. In fact, only recently \cite{Roca:2008} such a general
conclusion was presented for homogeneous degree networks, the case
of heterogeneous ones being well understood from other recent works
\cite{Santos:2006a,Gardenes:2007}.

In this paper we aim at extending our work on games on spatial structures and
homogeneous degree networks to the case when the updating of the strategies
follows the myopic best response rule \cite{Matsui:1992,Blume:1993}. There are a
number of reasons that support the relevance of such a study. First, previous
works on games on networks considered in general only imitative rules, i.e., a
specific individual updates her strategy by imitating the strategy of one of
her neighbors selected through different protocols (see e.g. \cite{Szabo:2007}
for a review). Such updating procedure makes only modest requirements on the
cognitive capabilities and/or information or memory
of the players: in these contexts, best response schemes
are the next step of sophistication, positing that individuals revise their
strategies by choosing the best reply to the strategies used by their neighbors
in the previous time step. This choice of updating based only on the previous
action of the neighbors is the reason why this dynamics is usually referred to
as myopic \cite{Ellison:1993}, although for brevity we will just use the term
``best response'' in what follows. Second, best response not only endows the
individuals of the model with more complete intellectual capabilities but also
is an innovative rule, as it allows extinct strategies to be reintroduced in
the system whereas imitative dynamics cannot do that. Third, best response is
the rule of choice in most studies from the economical viewpoint, the reason it
is not so often considered among physicists being that it gives rise to a
differential inclusion rather than a differential equation \cite{Hofbauer:1998}
and, subsequently, it is less amenable to analytical approaches. Finally,
earlier works on best response dynamics on lattices \cite{Blume:1993} left
interesting, hitherto unanswered questions such as the strong dependence of the
outcome of the evolution in coordination games on the initial conditions, a
point that we will specifically address here.

We discuss our results according to the following scheme: In Sec.\
\ref{sec:general} we summarize what is known about evolutionary games on
networks of homogeneous degree, which includes spatially structured populations
and random networks. We will consider in Sec.\ \ref{subsec:SD} the specific
case of the Snowdrift game \cite{Sugden:2004} as a paradigmatic example of the
difficulties arising in these studies. We will then proceed in Sec.\
\ref{sec:br} to present our simulation results on best response dynamics on
different types of networks and to subsequently discuss in detail the behavior
observed in lattices, providing an explanation as to why there may be more
or less cooperation on lattices than on the well-mixed case,
depending on the initial conditions. Finally, Sec.\
\ref{sec:conclusions} concludes the paper.

\section{Evolutionary games on networks of homogeneous degree}
\label{sec:general}

\subsection{2$\times$2 evolutionary games}
\label{subsec:evo}

The basic ingredient in the models we are going to discuss is evolutionary
game theory and, in particular, 2$\times$2 games.
Let us now briefly introduce the main concepts.
A symmetric $2 \times 2$ game
is  a game with 2 players who choose between 2 strategies and with no difference in role. Each player obtains a
payoff given by the following matrix
\begin{equation}
\label{eq:payoff-matrix}
\begin{array}{cc}
  & \begin{array} {cc} \mbox{C} & \mbox{D} \end{array} \\
  \begin{array}{c} \mbox{C} \\ \mbox{D} \end{array} &
  \left( \begin{array}{cc} 1 & S \\ T & 0 \end{array} \right).
\end{array}
\end{equation}
The rows represent the strategy of the player who obtains
the payoff and the columns that of her opponent.

The strategies are labeled as C and D for cooperate and defect,
because we interpret the game as a social dilemma. Indeed, certain
values of $S$ and $T$ undermine a hypothetical situation of mutual
cooperation. If $S<0$ a cooperator faces the risk of losing if the
other player defects, performing worse than with mutual defection. If
$T>1$ a cooperator has the temptation to defect and obtain a payoff
larger than that of mutual cooperation. Both tensions determine the
social dilemmas represented by symmetric $2 \times 2$ games
\cite{Macy:2002}. Restricting the values of the coefficients within
the intervals $-1 < S < 1$ and $0 < T < 2$, we have the Harmony game
\cite{Licht:1999} (HG, $0 < S,\, T < 1$) and three classic social
dilemmas: the Prisoner's Dilemma (PD, $-1 < S <
0,\; 1 < T < 2$), the Stag-Hunt game \cite{Skyrms:2003} (SH, $-1 < S
< 0 < T < 1$), and the Hawk-Dove \cite{Maynard-Smith:1973} or
Snowdrift game \cite{Sugden:2004} (SD, $0 < S < 1 < T < 2$). Each
game corresponds, thus, to a quadrant in the $ST$-plane.

To study the competition between cooperation and defection from an
evolutionary perspective, the payoffs obtained by playing the game
are considered as fitness and a darwinian dynamics is introduced to
promote the fittest strategy. The classic framework to do so is the
replicator dynamics \cite{Hofbauer:1998,Gintis:2000}, which assumes
an infinite and well-mixed population, i.e. a population with no
structure, where each individual plays with every other. Let $x$ be
the density of cooperators, and $f_c$ and $f_d$ the fitness of a
cooperator and a defector, respectively. The replicator dynamics
states that $x$ evolves as \cite{Hofbauer:1998}
\begin{equation}
\label{eq:repldyn-wellmixed}
\dot{x} = x (1-x) ( f_c - f_d ) .
\end{equation}
Then, if cooperators are doing better than defectors their density rises
accordingly, and the opposite occurs if they are doing worse. Provided that the
initial density of cooperators $x^0$ is different from 0 and 1, the asymptotic
state of this dynamical system is, for each game ($x^*$ represents the
asymptotic density of cooperators) \cite{Hofbauer:1998}: HG, full cooperation,
$x^* = 1$; PD, full defection, $x^* = 0$; SH, full cooperation if $x^0 > x_e$,
or full defection if $x^0 < x_e$; SD, mixed population with $x^* = x_e$,
regardless of the initial density $x^0$. Both in SH and SD the coexistence
equilibrium has a cooperation density $x_e = S / ( S+T-1 )$. It is
important to note that the outcome of these four games encompasses all the
possible cases for any symmetric $2 \times 2$ game \cite{Rapoport:1966} (see
also \cite{Roca:2006}).

\subsection{2$\times$2 evolutionary games on networks}
\label{subsec:nets}

As we stated in Sec.\ \ref{intro}, in 1992 Nowak and May \cite{Nowak:1992}
introduced spatial structure in the context of evolutionary games by
considering the players located at the nodes of a square lattice, playing the
game only with their neighbors (and playing the same action vs every one
of them) and not with the whole population. They
introduced evolution in this setup by using the unconditional imitation rule
(also known as ``imitate-the-best'' \cite{Szabo:2007}), where each player
chooses the strategy of the neighbor with largest payoff, provided this payoff
is greater than the player's. With this rule, they found that cooperators
survived by self-organizing in clusters, where the interactions within the
clusters yielded larger payoffs to cooperators than those
obtained by defectors at the boundaries of
cooperators' clusters.
\begin{figure*}
\begin{center}
\includegraphics[width=\textwidth]{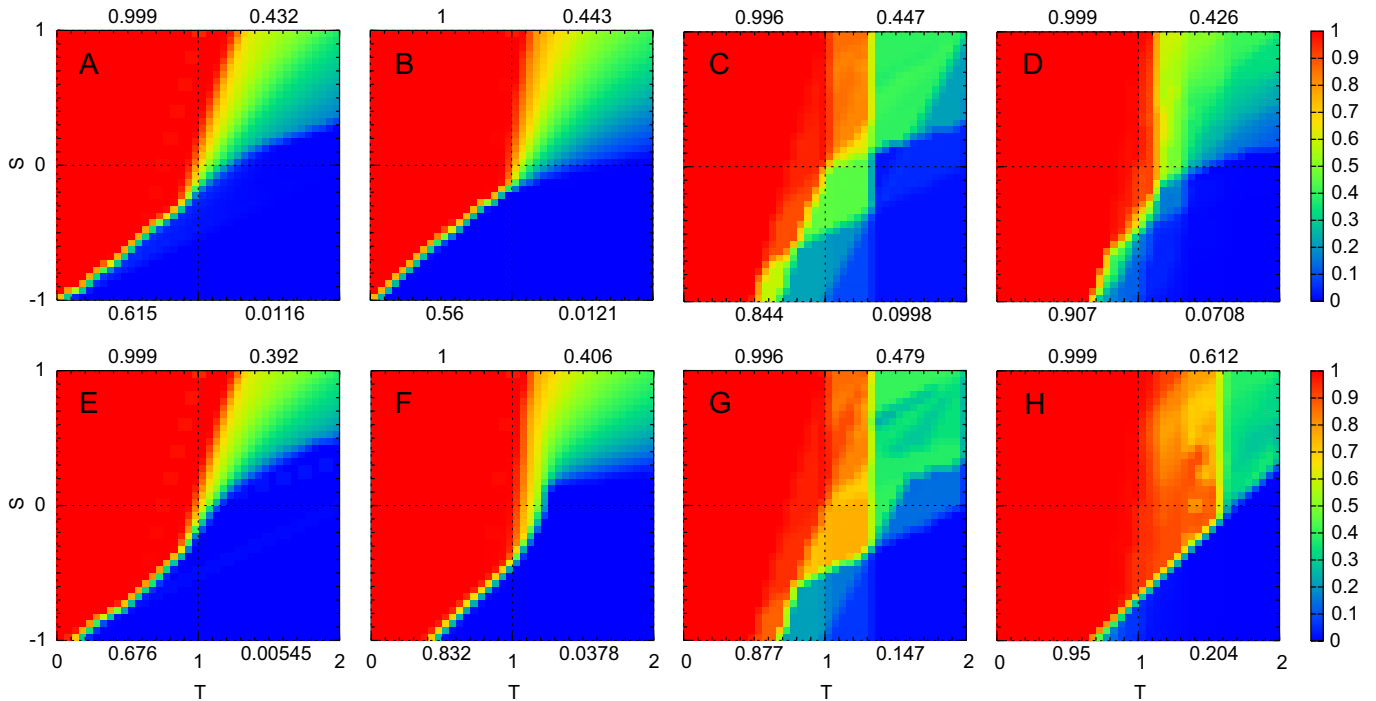}
\caption{\label{fig:1} Asymptotic density of cooperators $x^*$ in homogeneous random networks (upper row, A to D) compared to regular lattices (lower row, E to H), with degree $k=4$ (first and third columns, A, E, C, G) and $k=8$ (second and forth columns, B, F, D, H). Left plots (A, B, E, F) correspond to the replicator update rule, where right ones (C, D, G, H) use unconditional imitation as update rule(see text). The initial density of cooperators is $x^0 =0.5$ in all cases. The graphs display the key role of both the clustering of the network and the update rule (see main text). The promotion of cooperation is, in general, restricted to SH. The influence on PD is only significant when the update rule is unconditional imitation.}
\end{center}
\end{figure*}

Nowak and May's pioneering work opened the way to a large number of studies
focused on different games, different evolutionary rules, and different
lattices or networks. As a loose conclusion of those works, it was generally
believed that the existence of structure in the population, whether
spatial or of another kind, enhanced the emergence of cooperation in
games where defection was the norm. However, such conclusion did not agree with
all the available research, contradictions arose at several points (an example
of which will be discussed in Sec.\ \ref{subsec:SD} below) and there were no
studies that identified parameter regions where one
could firmly establish it. Therefore, in \cite{Roca:2008} we carried out a very
ambitious simulation program, as a result of which we were able to reach some
unambiguous conclusions. While a full report of our results is available in
\cite{Roca:2008}, we find it convenient to briefly recall here a few of the
ideas presented there, both as background information for the reader as well as
to introduce the way we will present our new results.

One of the most relevant findings reported in \cite{Roca:2008} was that the
spatial structure of a population, when modeled by a regular lattice, only has a significant effect on cooperation when the clustering coefficient is high, as seen by comparing results with those of homogeneous random networks of the same degree. This is illustrated by Fig.~\ref{fig:1}, where results for random homogeneous networks (upper row) are compared to results on regular lattices (lower row). Only when there is high transitivity or clustering in the network \cite{Newman:2003}, as occurs for regular lattices of degree $k=8$, significant differences appear.

Fig.~\ref{fig:1} also highlights the crucial influence of the update rule. Right columns show the results with unconditional imitation (introduced above), while the left ones present those obtained with the so-called replicator rule \cite{Hofbauer:1998,Gintis:2000}. This rule is defined as follows: Let $i = 1 \ldots N$ label the individuals in the population. Let $s_i$ be the strategy of player
$i$, $\pi_i$ her payoff and $N_i$ her neighborhood. With the replicator update
rule one neighbor $j$ of player $i$ is chosen at random, $j \in N_i$. The
probability of player $i$ adopting the strategy of player $j$ is given by
\begin{equation}
\label{eq:repldyn} p^t_{ij} \equiv \mathcal{P}\{
s_j^t \to s_i^{t+1} \} = \left\{ \begin{array}{l@{\quad:\quad}l}
  ( \pi_j^t - \pi_i^t ) / \Phi &ç
  \pi_j^t > \pi_i^t \\
  0 & \pi_j^t \leq \pi_i^t
\end{array} \right. ,
\end{equation}
 with $\Phi = k ( \max(1,T) - \min(0,S) )$ to ensure $\mathcal{P}(\cdot)
\in [0,1]$.

Fig.~\ref{fig:1} presents the results for the space of $2\times2$ games as a whole, using a color code which will be the same hereafter. Furthermore, we have introduced a quantitative measure $\mathcal{C}_G$ for the
overall asymptotic cooperation in game $G$ (= HG, PD, SH, SD), given by the
mean value of $x^*$ over the corresponding region in the $ST$-plane. This
global index of cooperation has a range $\mathcal{C}_G \in [0,1]$ and appears
on the graphs by the quadrant of each game. Both the qualitative assessment of the plots and the comparison of the values of this quantitative index yield an important regularity in the effect of the social structure modeled by this kind of networks: cooperation is generally enforced in coordination games (SH), specially when the clustering coefficient is high, whereas it is inhibited in anti-coordination games (SD). Remarkably, the positive effect on PD requires a particular update rule, namely unconditional imitation. We refer the interested reader to \cite{Roca:2008} for a complete discussion on these and related issues.

\begin{figure}
\begin{center}
\includegraphics[width=0.49\textwidth]{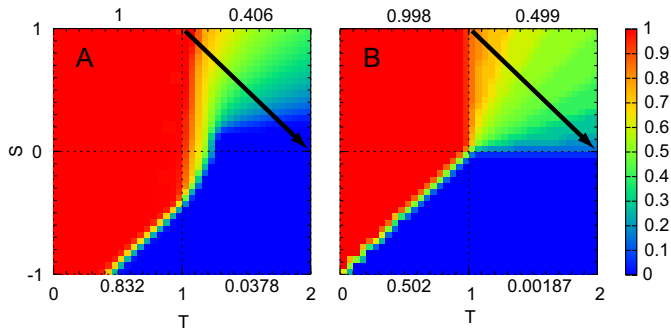}
\end{center}
\caption{Simulations of the models discussed in \cite{Hauert:2004} (A) and \cite{Sysi-Aho:2005} (B).
The lines mark the range of parameters studied in both works.
See text for a discussion.}
\label{fig:2}
\end{figure}

\section{The case of SD}
\label{subsec:SD}

Previous section introduced an important property of these evolutionary models, the crucial dependence that the result has on the update rule. We will now discuss a specific example which, on the one hand, allows us to make this point and, on the other hand, will introduce us to our main subject,
namely the effects of best response dynamics.

In 2004, Hauert and Doebeli \cite{Hauert:2004} presented a study of the
Snowdrift game in which they concluded that, contrary to the general belief,
the existence of a spatial structure may inhibit cooperation. To that end,
they studied the dependence of the level of cooperation on the parameter
$r$, the cost-benefit ratio (understood as cost of cooperating and
corresponding benefit accrued), which in our parameterization is $r=(T-S)/2$.
They did not consider the whole $ST$-plane but only the line given by
$T=1+r$, $S=1-r$ (see Fig.\ \ref{fig:2}) and obtained asymptotic levels
of cooperation below those found in well mixed populations.

One year later, Sysi-Aho and coworkers \cite{Sysi-Aho:2005} repeated the same
study changing only the dynamical rule: Where Hauert and Doebeli had used the
replicator rule mentioned above, Sysi-Aho and coworkers used the best response
rule, introducing a probability $p<1$ to update strategy, i.e., at every time
step every player chose her strategy as a best response to their neighbors with
probability $p$ or left it unchanged with probability $1-p$. The reason to do
that is to prevent the systems from falling onto a sequence of alternate states
of full defection and full cooperation, which is an artifact of the rule (and
which are never reached as soon as $p<1$, see \cite{Sysi-Aho:2005}). They
studied the same range of parameters, finding that cooperation subsisted even
for $r$ close to 1, being larger (resp.\ smaller) than in a well mixed
population for large (resp.\ small) $r$.

\begin{figure*}
\begin{center}
\includegraphics[width=0.75\textwidth]{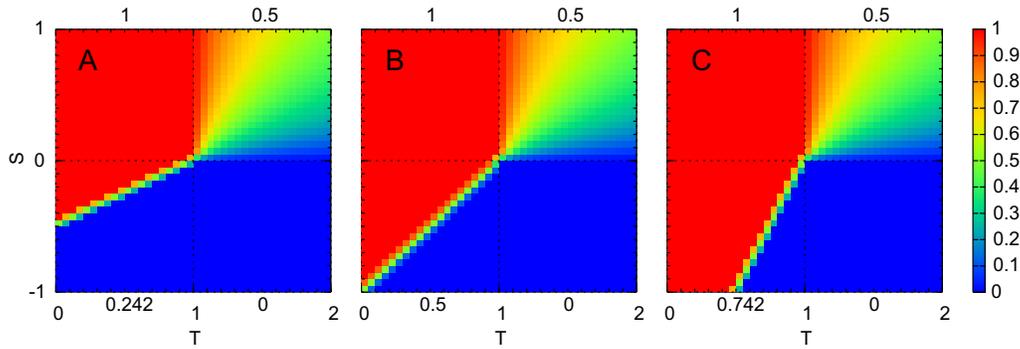}
\end{center}
\caption{
\label{fig:3}
Asymptotic density of cooperators $x^*$ in well mixed populations under best response dynamics, with $p=0.1$ (see main text), when the initial density of cooperators is $x^0 =1/3$ (left, A), $x^0 =1/2$ (middle, B) and $x^0 =2/3$ (right, C).}
\end{figure*}

In order to understand better this issue, we have reproduced the simulations
in \cite{Hauert:2004,Sysi-Aho:2005}, using the same square lattices, neighborhoods
($k=8$) and initial conditions (cooperation and defection equally likely)
they used, obtaining the results we summarize in Fig.\
\ref{fig:2}. As we may see, the two different dynamics lead to rather different
results when looked at in the framework of the whole $ST$-plane. We stress that,
along the line indicated in the plots, we exactly reproduce the results reported
in those previous studies. There is indeed a decrease of cooperation when
using the replicator dynamics as in \cite{Hauert:2004}, along the line of interest
but also in the SD quadrant as a whole. However, when the dynamics is best
response we do observe a stepped profile in the SD quadrant as Sysi-Aho {\em
et al.} did, while the mean value of the cooperation over the quadrant
is the same as in a well mixed population. The
reader is referred to the middle panel of Fig.\ \ref{fig:3}, where the results on
a well mixed population with the same initial condition are depicted.

These results open up a series of questions, beginning with the following: What
is the effect of best response dynamics on other networks, given that on a
square lattice its effect is not very noticeable? Let us recall that best
response is a step further towards ``intelligence'' of the agents as compared
to the replicator rule, an imitative, non-innovative dynamics, and therefore we
might expect that players could exploit better the existence of a network. This
issue is what we discuss in detail in what follows. However, there is a more
general problem, namely what does it mean ``promotion of cooperation by
the structure of a population''? Do we refer to a specific set of
parameters, such as the line studied in \cite{Hauert:2004,Sysi-Aho:2005}? Do we
refer to a global measure of the cooperation level in a region, such as the
values we compute for each quadrant? Or do we refer to the  $ST$-plane as a
whole? Note that in this last case the replicator rule of Hauert and Doebeli,
while indeed leading to less cooperation in the SD game, yields a very large
increase of the cooperative region in the SH quadrant, favoring players to
coordinate in the Pareto-dominant equilibrium \cite{Gintis:2000}. Had we been
looking at that quadrant only, we would have certainly concluded that
cooperation is promoted by the square lattice. Or, looking at the SD
game, had we considered unconditional imitation as the update rule, we would
have found evidence of spatial structure fostering cooperation in SD. While we
will not dwell any further in this issue here (but see
\cite{Roca:2008,Roca:2009} for a discussion in depth of these problems) we want
to stress that statements about promotion of cooperation should be made in a
much more specific manner without trying to attach to them unchecked general
implications beyond the scope of the case under study.

\section{Best response dynamics}
\label{sec:br}

\subsection{Different types of networks}

\begin{figure*}
\begin{center}
\includegraphics[width=0.75\textwidth]{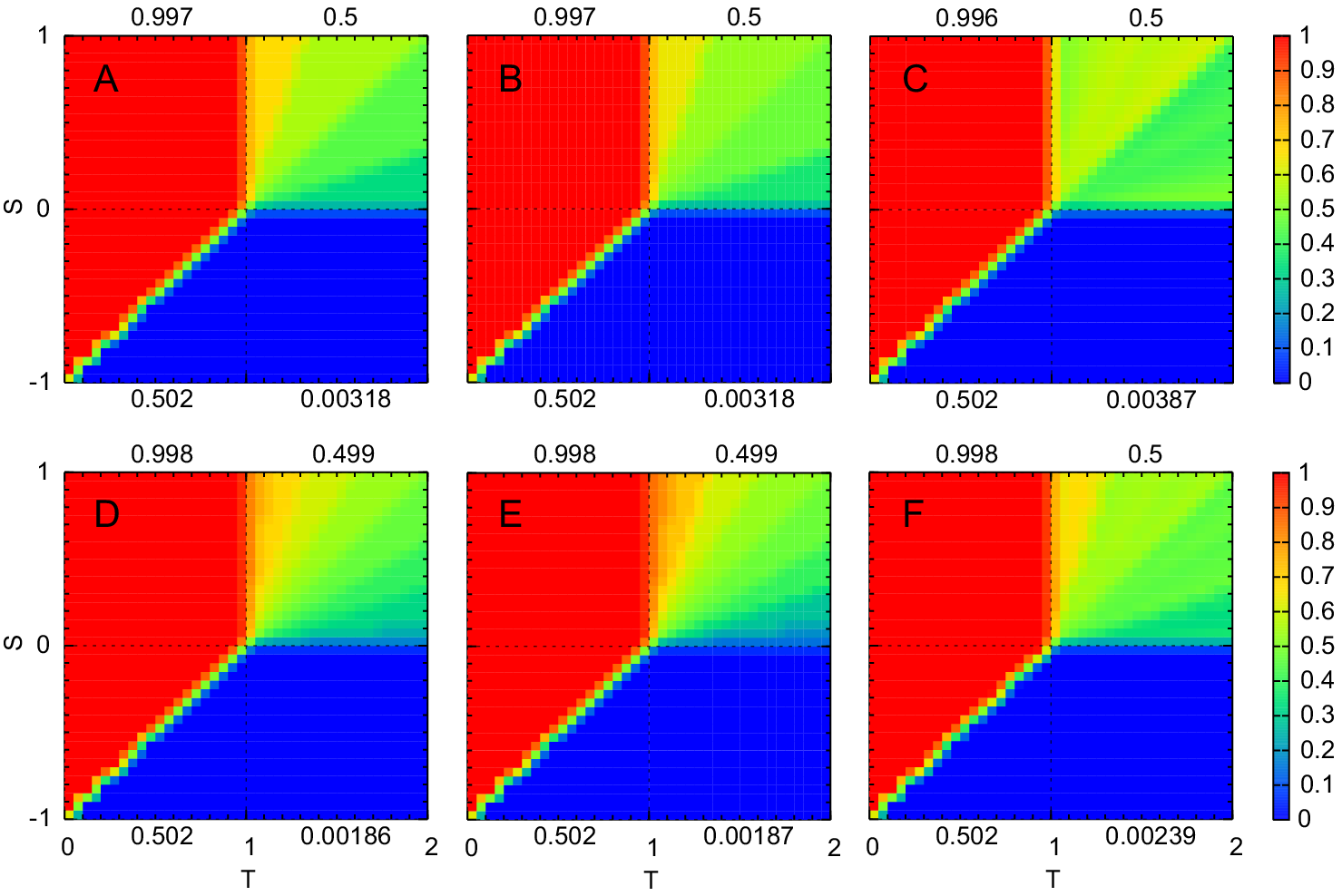}
\end{center}
\caption{Asymptotic density
of cooperators $x^*$ in random (left, A and D), regular (middle, B and E), and scale-free networks (right, C and F) with degrees $k=4$ (upper row, A to C)
and 8 (lower row, D to F). The update rule is best response with $p=0.1$
and the initial density of cooperators is $x^0 =0.5$.
Differences are negligible in all cases; note, however, that the steps appearing in the SD quadrant are slightly different.}
\label{fig:4}
\end{figure*}

\begin{figure}
\begin{center}
\includegraphics[width=0.49\textwidth]{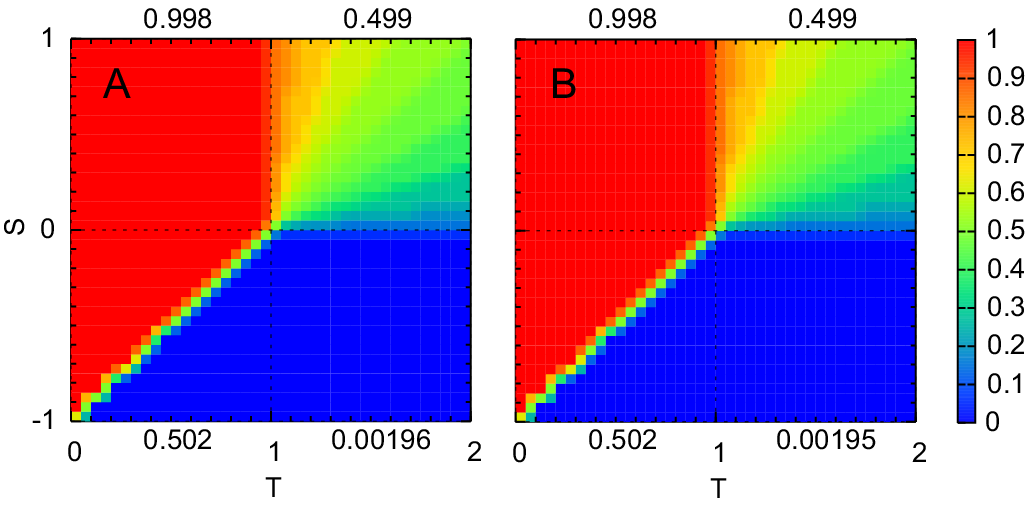}
\end{center}
\caption{\label{fig:5}
Asymptotic density of cooperators $x^*$ in regular lattices
with degree $k=8$. The update rule is best response and the initial density of cooperators is $x^0 =0.5$. Left (A): asynchronous updating with $p=0.1$; right (B): synchronous updating with $p=0.01$. Both results are virtually identical to that of Fig.~\ref{fig:4}~E, which was obtained with synchronous updating and $p=0.1$.}
\end{figure}

Motivated by the reasons discussed in the introduction and by the issues raised
by our study in Sec.\ \ref{subsec:SD}, we undertook the study of the effects of the
best response dynamics on a large family of networks. We considered lattices,
which may represent spatial structure, homogeneous random networks (random
networks where all nodes have exactly the same degree),
Erd\"os-Renyi random networks \cite{Bollobas:2001}, small-world networks, Barab\'asi-Albert
\cite{Barabasi:1999}
scale free networks and Klemm-Egu\'\i luz \cite{Klemm:2002} scale free networks
with different mean degrees, thus exploring all possible combinations of small-world
phenomena, scale free behavior and large or small clustering. We have also
considered the complete graph as the reference for a well mixed population,
results for which are presented in Fig.\ \ref{fig:3}.

We will begin by focusing on the case in which the initial conditions are a
50\% of cooperators and a 50\% of defectors. Our reference for comparison will
then be Fig.~\ref{fig:3}~B, corresponding to a well-mixed (complete network)
population with that initial condition. A small subset of our results for the
other networks is presented in Fig.~\ref{fig:4}, showing that the asymptotic
behavior does not depend at all on the type of network considered, with the only and unimportant exception of slight differences in the stepping in SD.
We want to stress that we have tested many other networks aside from those presented here with exactly the same results. Not only the mean cooperation levels per
quadrant are practically the same, but also the dependence on $S$ and $T$ for
each quadrant. We note that this is a very remarkable result, in so far as for
most other (imitative) dynamics studied there is always a largely noticeable
effect of the type of network on which the games are played, as we have seen
above (cf.\ Figs.\ \ref{fig:1} and \ref{fig:2}; see also
\cite{Roca:2008,Santos:2006a} for more details on homogeneous and scale free
networks, respectively). {We have made every effort to ensure that
our results are robust and independent of the technicalities of the
simulations. To begin with, we have checked a correct convergence, verifying that
simulation times larger up to a factor of 10 lead to the same results}. On the
other hand, the probability parameter $p$, which governs how often agents
update their strategy, is not relevant, and neither it is the use of asynchronous updating (see Fig~\ref{fig:5}), in which an agent is chosen at random to update its strategy after a round of the game has been played by her and her neighbors (we note that asynchronicity was reported to have a small but noticeable effect with imitative update rules in
\cite{Roca:2008}). With all these checks, we can finally state our first
conclusion, namely that best response dynamics is insensitive to the type of
network on which the game is played, meaning that when the networks are
initiated from the same proportion of cooperators and defectors the asymptotic
results are the same as well. This unexpected result may have interesting
implications in applied contexts, implications which we will discuss in the
concluding Section.

\begin{figure*}
\begin{center}
\includegraphics[width=0.75\textwidth]{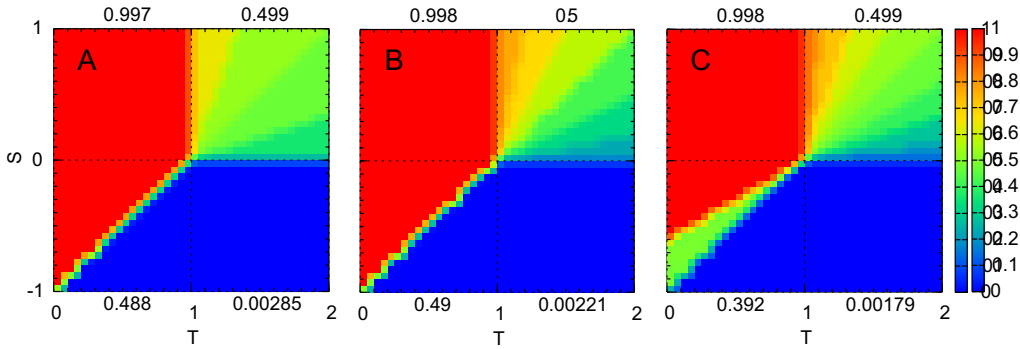}
\end{center}
\caption{Asymptotic density
of cooperators $x^*$ in regular  lattices with initial density of cooperators $x^0 =1/3$. The degrees are $k=4$ (left, A), $k=6$ (middle, B) and $k=8$ (right, C). The update rule is best response with $p=0.1$.
Equivalent (symmetrical) results are obtained for $x^0 =2/3$ (see main text).}
\label{fig:6}
\end{figure*}

\subsection{The effect of initial conditions}
\label{sec:lattices}

Having found that best response is largely unaffected by playing the different
2$\times 2$ games on a network, we extended our analysis of the problem to
consider different initial conditions. We initiated our simulations with an
initial density of cooperators $x^0 =1/3$ or $x^0 =2/3$ and repeated our
simulation program for our family of networks. In order to understand the
observed phenomenology, we find it convenient to begin the discussion by the
case of lattices. Figure \ref{fig:6} shows the results for lattices with
different number of neighbors and $x^0 =1/3$ (the outcome of the simulations
for $x^0 =2/3$ is similar, with the green region in the SH quadrant of Fig.\
\ref{fig:5} being symmetrically located below the $S=T-1$ line). As can be seen from Fig.\ \ref{fig:6}, for 4 and 6 neighbors the results for $x^0=1/3$ are
indistinguishable from the results for $x^0=1/2$. This must be compared with
Fig.~\ref{fig:3}~A, which shows for a well-mixed population a cooperative region in the SH quadrant that is roughly a 50\% of the one we have obtained
on lattices. As the asymptotics is the same for lattices with 4 and 6 neighbors, we also conclude that the clustering of the network does not play any role. On the other hand, the lattice with 8 neighbors shows a striking result, namely the
appearance on the SH quadrant of a region with intermediate values of
cooperation. Let us recall that, as a stand-alone game, SH has two equilibria,
full cooperation or full defection, and that depending on the initial condition the system ends up in one or the other. This is the behavior observed in all
the lattices and networks studied so far with imitative rules and also was what we observed for best response with $x^0=0.5$.

\begin{figure*}
\begin{center}
\includegraphics[width=\textwidth]{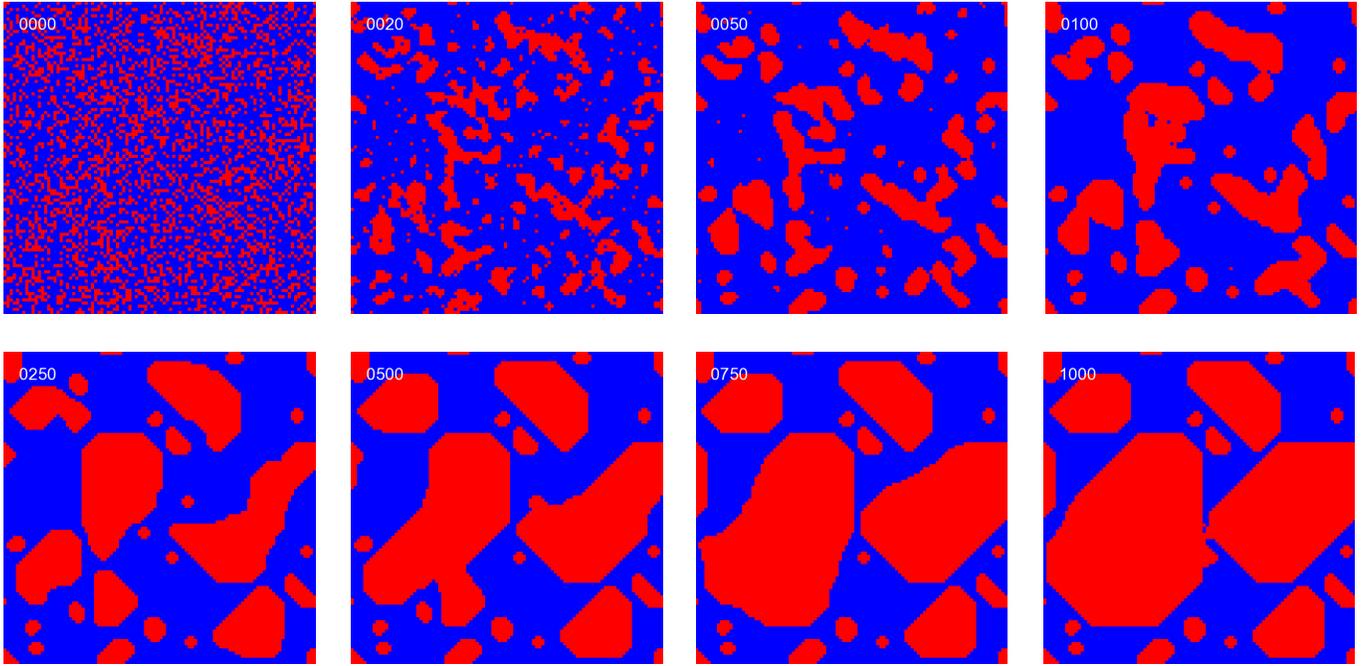}
\end{center}
\caption{Snapshots of the evolution on a regular lattice of degree $k=8$, under best response dynamics with $p=0.1$. Sites depicted in red are cooperators, blue ones are defectors. The initial density of cooperators is $x^0 =1/3$. The game parameters are $S=-0.6$, $T=0.2$, in the SH quadrant. Times are as indicated in the snapshots. System size is 100$\times$100 with periodic boundary conditions. In this particular realization, cooperation eventually dominates entirely the lattice.}
\label{fig:7}
\end{figure*}

In order to understand this surprising feature, the appearance of regions
with seemingly mixed behavior in SH and its dependence on the number of
neighbors, appearing for 8 neighbors but not for smaller numbers, we have
looked in detail at the time evolution of specific realizations. Fig.~\ref{fig:7} presents a representative example of this dynamics with 8 neighbors and
for values of $S$ and $T$ in the intermediate region we are interested in.
We clearly observe that there is an initial stage in which isolated cooperators
(in red) die out and only clusters with a significantly larger presence of
cooperators avoid extinction and eventually begin to grow. A careful look
at the pictures and at the whole time evolution allows one to notice that
clusters grow due to the advancement of some of its sides, in particular
those which are basically flat but have just one kink, whereas some others
are stopped unless hit by some other advancing front. Very clearly observable
among the latter are diagonal fronts (see, in particular, the last three frames
in the sequence). This can be easily understood analytically by just considering
the different possible types of front and the conditions for their advancement in
terms of best response dynamics. Thus, for a planar front such as
\begin{center}
\sf
\begin{tabular}{cccccc}
C & C & C & D & D& D \\
C & C & C & D & D& D \\
C & C & C & D & D& D \\
C & C & C & D & D& D \\
C & C & C & D & D& D \\
C & C & C & D & D& D
\end{tabular}
\end{center}
the borderline defectors
will become cooperators if $S>3(T-1)/5$, whereas cooperators are
transformed into defectors if $S<5(T-1)/3$. Therefore, in the intermediate
region $5(T-1)/3<S<3(T-1)/5$ planar fronts are stable in the lattice with
8 neighbors. It is easy to check that diagonal fronts are stable in this
region as well. This must be compared with the situation when there is a
kink in the front, as in
\begin{center}
\sf
\begin{tabular}{cccccc}
C & C & C & D & D& D \\
C & C & C & D & D& D \\
C & C & C & D & D& D \\
C & C & C & C & D& D \\
C & C & C & C & D& D \\
C & C & C & C & D& D
\end{tabular}
\end{center}
In this case the defector at the kink becomes a cooperator if $S>T-1$, and the
cooperator at the kink becomes a defector in the opposite case, $S<T-1$.
Therefore, for this type of front there is no stability region and it always
advances in one direction or the other. This analysis shows that the fate of a
specific realization depends on the geometry of the clusters arising at its
first stage. What we see in the special region in Fig.~\ref{fig:6}~C is the
result of some simulations that cannot reach full cooperation because the
growing clusters arrest at some point in the simulations, whereas in other
cases their geometry is such (as in Fig.~\ref{fig:7}) that the system ends
up dominated entirely by cooperators. We note that the discussion is
symmetrical with respect to the initial condition and it can be applied to
explain the results for $x^0=2/3$ (not shown).

It remains to be explained why the other lattices do not show this
region of intermediate behavior. The reason can again be traced
back to the geometry of the clusters: Consider, for instance, the case
with 4 neighbors. One can then show that diagonal fronts are never
stable and, in fact, are subject to the same conditions as planar
fronts with a kink, making it much more difficult for a particular
realization to have all its clusters arrested. Hence, in the simulations on the lattice with 4 neighbors situations with a mixed population of cooperators and defectors are never found. A similar reasoning explains the results for the 6 neighbor lattice.

\begin{figure}
\mbox{ }\\[2mm]
\begin{center}
\includegraphics[width=0.49\textwidth]{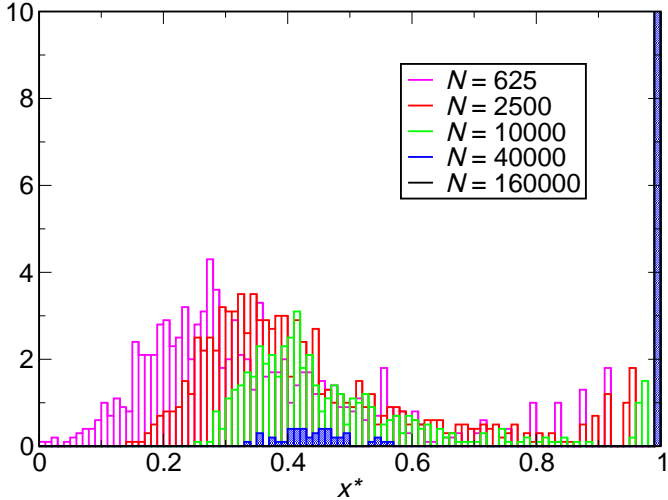}
\end{center}
\caption{Normalized histogram of the asymptotic densities of cooperators $x^*$ reached in the simulations on regular lattices of degree $k=8$ with initial density of cooperators $x^0 =1/3$. The update rule is best response with $p=0.1$. Sizes are as indicated in the plot. Note that the value at $x^*=1$ goes out of range for the two largest sizes. $S$ and $T$ are as in Fig.~\ref{fig:7}.}
\label{fig:8}
\end{figure}

The mechanism behind the appearance of the mixed populations suggests that the
coexistence of cooperators and defectors in the SH may be due to finite size
effects. To check this possibility, we have simulated the lattice with 8
neighbors on a large variety of sizes, plotting a histogram of the number of realizations that stop on a mixed population. The result is depicted in Fig.\ \ref{fig:8}, showing clearly that as the system size grows the number of asymptotically
mixed realizations becomes negligible. This is a strong indication that in the
infinite size limit the 8 neighbor lattice will behave as those with less
neighbors, arriving always at the same asymptotics as for the $x^0=0.5$ initial
condition, at least for initial data not too close to $x^0=0$ or $x^0=1$. We
note, however, that for applications of these ideas to real life problems,
where populations (e.g., in a social context) are finite, it may be possible to
observe mixed populations in SH dilemmas under best response dynamics.

\begin{figure}
\begin{center}\includegraphics[width=0.49\textwidth]{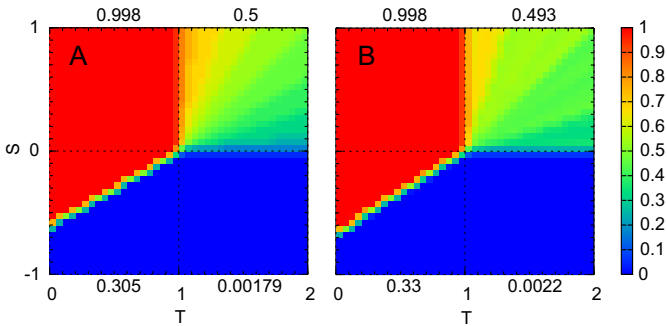}
\end{center}
\caption{Asymptotic density
of cooperators $x^*$ in homogeneous random (left, A) and scale free (right, B)
networks with initial density of cooperators $x^0 =1/3$.
The update rule is best response with $p=0.1$.
Equivalent (symmetrical) results are obtained for $x^0 =2/3$.}
\label{fig:9}
\end{figure}

Finally, with the understanding of the observations on lattices at hand,
we have also looked at the effect of initial conditions on other networks.
Two examples of our simulations are presented in Fig.~\ref{fig:9}.
By comparing with the well mixed result in Fig.~\ref{fig:3}~A, we see that
initial conditions {different from $x^0=0.5$ also give rise to non
trivial effects} on random and scale free lattices. No regions of mixed
populations are observed, and the mean cooperation level increases slightly
when going from well mixed to homogeneous random and from there to scale free
networks, {but always to a lesser extent than on lattices}.
Unfortunately. the topology of these networks renders the analysis in terms of
clusters unapplicable. We do not have a clear understanding of this small
promotion of cooperation (inhibition in the case $x^0=2/3$) on these networks,
{but we can tentatively hypothesize the development of weak, long
range correlations, which would be responsible for such a weaker effect, in
contrast to those of lattices, of a short and stronger nature}.

\section{Discussion and conclusions}
\label{sec:conclusions}

In this paper we have presented a thorough analysis of myopic best response
dynamics on $2\times 2$ social dilemmas with different types of networks. The
original motivation of our results was the clarification of the discrepancies
among the work of Hauert and Doebeli \cite{Hauert:2004} and Sysi-Aho {\em et
al.} \cite{Sysi-Aho:2005}, who reported an inhibition (resp. promotion) of
cooperation in the SD game on square lattices. We have shown that the problem
is that the comparison is basically meaningless in so far as they use different
update rules (proportional update or replicator dynamics vs best response) and
also because restricting the comparison to SD games leads to missing the larger
picture of the set of possible dilemmas. Indeed, in the proportional update
dynamics there is a large increase of cooperation in the SH quadrant, which
contrasts with the phenomenology of the SD dilemma (as is generically the case,
see \cite{Roca:2008}).

Our analysis of myopic best response dynamics has yielded two main results:
First, there is practically no effect of the type of network considered on
the asymptotic behavior for any value of the game parameters; and,
second, there is a noticeable effect of the initial conditions with particular
importance in the case of lattices. The first result is most interesting,
more so if viewed in the context of previous work on imitative rules
\cite{Roca:2008,Santos:2006a}. In principle, given the definition of best
response dynamics, one would not expect important effects on the
Harmony and PD quadrants, where there is only one global best
response and therefore the network can not introduce any novel
feature. However, the fact that SD has only one stable equilibrium of
a mixed character, that may be difficult to fulfill in the presence of a
network, or the bistable nature of SH allow to expect some influence
from the interaction networks. The fact that there is none (at least at a
global level, for the whole network; local peculiarities are not being
considered here) is therefore quite remarkable and casts a shadow of
doubt on the applicability of the studies of imitative rules to specific
socio-economic contexts. Indeed, if using a rule with a higher degree
of intelligence renders the network effect unnoticeable, whereas it is
crucial for simple (non-innovative) dynamics, one can presume that
the behavior of actual economic agents will be closer to best response
and hence network-independent to some extent. Interestingly, this
idea may be related to the fact that models of network formation in
economics often lead to very simple networks (see, e.g.,
\cite{Jackson:1996,Goyal:2007}, and references in the latter),
which might be another hint of the secondary role played by the
network structure in economical applications.

On the other hand, we have also seen that there are network effects even under
best response dynamics, that arise in the dependence of the asymptotic behavior
on the initial conditions. Lattices are prominent examples of these phenomena
and, using the regularity of their structure, we have been able to trace the
differences with well-mixed populations to the formation of clusters, whose
stability depends on the degree of the lattice (at least for finite size
lattices). We have thus seen that cooperation may be promoted (resp. inhibited)
in the SH quadrant in situations when the initial amount of cooperators would
be insufficient (resp. sufficient) to take the system to full cooperation. The
effect is also observed on other networks, always with the same symmetrical
character, but the reason for this behavior remains still an open issue. Again,
in a economical context this result is interesting because it is generally
believed that best response does not exhibit any new feature when used on a
network, whereas here we see that the dependence on the initial conditions is
not the same as in the complete network.

Finally, from a more general viewpoint, we have presented yet another example
that the phenomenology of social dilemmas on networks is largely non universal.
While best response turns out to be peculiar in the sense that it gives rise to
network-independent behavior, it is important to realize that once again
changing the dynamics of the strategy update leads to large, non-trivial
changes in the results of evolution as compared to other rules. Furthermore, as
we have already mentioned, only a full study of the whole space of dilemmas
under consideration (in our case, $2\times 2$ games) can shed some light on the
mechanism governing the evolution on different networks and under different
rules. In this sense, we have presented here a {detailed} analysis
of best response games which highlights the fact that, as compared to what is
observed on well-mixed populations, the dilemma that is most affected is SH,
i.e., risky situations, rather than contexts in which the important tension is
the temptation to defect. This is clearly related to the bistable character of
SH, in which the best response {tends to be equal to} the opponent's
actions, and suggests that similar mechanisms may be at work in games and
dilemmas with more players and/or strategies. Further research along these
lines is needed to confirm these intuitions.

\section*{Acknowledgments}
We are grateful to Constanza Fosco for a critical reading of the manuscript.
This work is partially supported by Ministerio de Ciencia e Innovaci\'on
(Spain) under grants Ingenio\--MATHE\-MATICA and MOSAICO, by Comunidad de
Madrid (Spain) under Grants  MOSS\-NOHO-CM and  SIMUMAT-CM and by European
Science Foundation under COST Action P10.

\end{document}